\documentstyle[prl,epsf,aps]{revtex}

\begin{document}
\bibliographystyle{plain}
\title{From inherent structures to pure states:
some simple remarks and examples.}
\author{Giulio Biroli and R{\'e}mi Monasson}
\address{
Laboratoire de Physique Th{\'e}orique de l'Ecole Normale 
Sup{\'e}rieure\thanks{Unit{\'e} Mixte de Recherche du Centre National de la 
Recherche Scientifique et de l'Ecole Normale Sup{\'e}rieure.},
\\ 24 rue Lhomond, 75231 Paris cedex 05, France.}


\maketitle

\begin{abstract}
The notions of pure states and inherent structures, {\em i.e.}  stable
configurations against 1-spin flip are discussed.  We explain why
these different concepts accidentally coincide in mean-field models
with infinite connectivity and present an exactly solvable
one dimensional model where they do not. At zero temperature pure
states are to some extent related to $k$-spin flip stable
configurations with $k\to \infty$ after the thermodynamical limit has
been taken. This relationship is supported by an explicit analysis of
the TAP equations and calculation of the number of pure states and
$k$-spin flips stable configurations in a mean-field model with finite
couplings. Finally we discuss the relevance of the concepts
of pure states and inherent structures in finite dimensional glassy systems.
\end{abstract}

\section{Introduction}
In a series of seminal papers, Kirkpatrick, Thirumalai and Wolynes
(KTW) suggested ten years ago that (discontinuous) 
mean-field spin glasses could serve
as a paradigm for vitreous systems\cite{Ktw}.  The major success of
KTW theory lies in its ability to predict the existence of both low
(Kauzmann) and high (dynamical freezing)
temperature critical points and thus to describe in a unified
framework the thermodynamical and dynamical signatures of the glassy
transition\cite{Angell}. This achievement relies on a simple but rich
picture of glassy {\em pure states} (PS), that is local minima in the
{\em free-energy} landscape, and of the dynamical evolution taking
place in the latter. To what extent KTW mean-field picture applies to
real systems is a crucial issue\cite{first,second}.

An apparently related approach based on the investigation of the {\em
potential energy} landscape and of its local minima called {\em
inherent structures} (IS) had been already proposed at the beginning of the
eighties by Stillinger and Weber in the context of liquid
theory\cite{Stillinger}. IS present a considerable advantage with
respect to PS: they are well defined and free of any mean-field
hallmark. Recently, numerical studies proposing that some understanding
of the glass transition could be gained from the analysis of potential
energy landscapes\cite{sciort} strengthened the feeling that
Stillinger and Weber's analysis and KTW theory were basically the same
description of glassy systems. This point of view was later supported by the
equivalence of IS and PS in some mean-field spin-glasses, e.g.
Sherrington-Kirkpatrick (SK) or $p$-spins models\cite{felix} 
and by analytical works\cite{first,third}.

In this letter, we present some arguments to clarify the relationship
between IS and PS. We present some one imensional models for which IS
can be exactly computed and do not coincide with PS.  We show in
addition that the degeneracy between IS and PS in infinitely-connected
models is due to the vanishing of the interactions in the
thermodynamical limit and is lifted by the introduction of finite
couplings. Focusing on the TAP equations ``defining'' PS and the number of
the latters, we show that zero temperature PS are related to extended
IS, that is configurations stable with respect to an arbitrary large
number $k$ of spin flips (and not only to a single one as usual
IS)\cite{groundstate}.

\section{Definitions and notations}
We consider a model including $N$ Ising spins $S_i$. A configuration
of spins will be said $k$-stable if its energy cannot be decreased by
flipping any subset of $k$ (or less than $k$) spins. Let us call $s_k
(e)$ the logarithm (divided by $N$) of the number of $k$-stable
configurations with energy density excess $e-e_{GS}$ with respect to the
ground state energy density $e_{GS}$ when $N\to \infty$. IS correspond
to 1-stable configurations and $s_1(e)$ is usually called
configurational entropy. We have of course $s_{k+1} (e) \le s_k (e)$
and $s_k (e) =-\infty $ for $e<e_{GS}$.

The notion of PS is much trickier.  In mean-field
systems PS are usually defined through TAP equations \cite{tap} for the local
magnetisations $m_i$. TAP solutions are exponentially numerous in the
volume $N$ and can be accounted for through the so-called complexity
$\sigma (f)$, that is the normalised logarithm of PS having
free-energy densities $f$.  We shall focus in the following on the
zero temperature limit of the complexity $\sigma (e)$ and compare it
to $s_k (e)$.

We shall denote hereafter the maximum over $e$ of $s_k (e)$
(respectively $\sigma (e)$) by $s_k$ (resp. $\sigma$).

\section{One dimensional disordered chain}
We first consider a Ising chain with disordered nearest neighbour couplings
$L_i$ and no external field\cite{Luck}. The $L_{i}$s are 
independent random variables drawn from an even distribution
$P(L)$. No phase transition can take place at finite temperature:
there is only one paramagnetic PS\cite{Luck}. Therefore, the zero
temperature limit of the complexity simply reads $\sigma (e) = 0$ if
$e=e_{GS}$, $\sigma (e) =-\infty$ if $e \ne e_{GS}$
where the ground state energy density equals $e_{GS} = -\int dL P(L) |L|$.

Ettelaie and Moore have calculated $s_1 (e)$\cite{Moore}. We extend
their approach to obtain $s_k(e)$ using the following observation by
Li: a necessary and sufficient condition for a configuration to be
$k$-stable is that each frustrated bond be $k$-weak, {\em i.e.}
smaller in magnitude than the $2k$ nearest bonds ($k$ on the left, $k$
on the right) on the chain\cite{Li}. The typical and self-averaging
$s_k (e)$ thus reads
\begin{equation}\label{sk1d}
s_k (e)=\lim _{N\to \infty} \frac 1N
\ln \left( \sum_{\{\tau _\ell = 0, 1\}} \delta \left( e - e_{GS}-\frac 2N 
\sum_{\ell=1}^{N_k}|L_\ell | \tau _\ell \right) \right)\qquad ,
\end{equation}      
where $N_k$ is the number of $k$-weak bonds on the chain and $\tau
_\ell$ equals 1 if the $\ell ^{th}$ $k$-weak bond is frustrated, 0
otherwise. Using an integral representation of the delta function
in (\ref{sk1d}), the $\tau_\ell$'s may be traced over and one finds
\begin{equation}\label{s(k,e)fin1}
s_k (e)=\lim _{N\to \infty} \frac 1N
\ln \left\{ \int_{-i\infty}^{+i\infty}\frac{du}{2\pi N i}
\exp N\left( u (e - e_{GS} ) - 2\int_{0}^{+\infty} dL \;
P_k(L ) \; \ln(1+e^{2Lu}) 
\right)\right\}  \ .\label{s(k,e)fin2}
\end{equation}
where $P_{k}(L)=P(L) (2\int_{|L|}^{\infty}dL'\;P(L') \;)^{2k}$ denotes
the probability that a bond is $k$-weak and has value $L$.  The
integral over $u$ can be evaluated by the method of steepest descent.
In fig.1, $s_k (e)$ is plotted for a Gaussian $P(L)$ (with variance
unity) and for different values of $k$. At fixed $k$, $s_k (e)$ is
symmetric around its maximum located in $e_k = e_{GS} + \int dL P_k(L)
|L|$, $s_k = \ln 2 /(2k+1)$. Note that the height of the maximum $s_k$,
related to the total number of $k$-stable configurations is
independent of $P$ as found by Li\cite{Li}. On the left boundary,
all curves exhibit an infinite slope when $e\to e_{GS}$. More
precisely, $s_k (e_{GS} + \varepsilon ) \propto
-\varepsilon\ln\varepsilon $ if the support of $P(L)$ does not contain 
$L=0$ and $s_k (e_{GS} + \varepsilon ) \propto
\varepsilon ^{(\alpha +1)/(\alpha +2)}$ if $P(L)\propto L^{\alpha}$
when $L\to 0$ (with $\alpha > -2$ to make $e_{GS}$ finite).

\begin{figure}[bt]
\centerline{    \epsfysize=6cm
       \epsffile{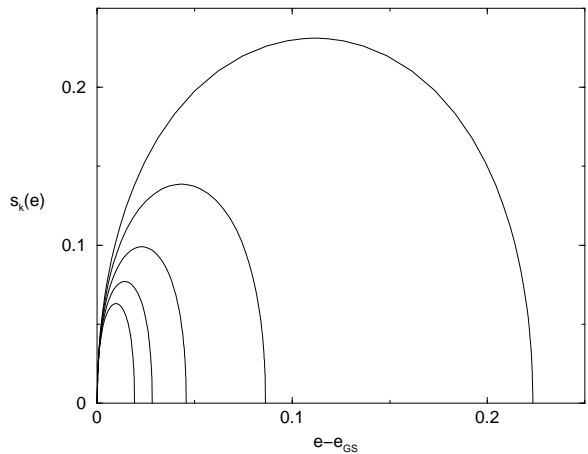}}
\caption{Logarithm of the number of $k$-stable configurations 
$s_{k}(e)$ as a function of the excess of energy density $e-e_{GS}$ 
for a Gaussian coupling distribution $P(L)$ (with variance 
unity and zero mean). From top to bottom: $k=1,2,3,4,5$.}
\end{figure}

The above model is a clear example of a system with many IS and at the
same time a single PS.  More precisely, the zero temperature limit of
the complexity $\sigma (e)$ (related to PS) is not equal to the
configurational entropy $s_1(e)$ counting IS but to $s_\infty
(e)$. Before exposing why $s_1$ and $\sigma$ accidentally coincide in
infinitely connected spin-glasses, let us briefly see to what extent
the above results apply to generic finite dimensional systems.

\section{General remarks for finite dimensional systems}
The main features of $s_k (e)$ can be found by some general arguments
which apply to any {\em finite dimensional} system. The behaviour of
$s_1 (e_{GS} + \varepsilon)$ at small $\varepsilon$ has been predicted
in \cite{logx1,logx2} and depends only on the lowest-lying excitations of the
system.  In addition, the scaling of the right-edge energy $2 e_k
-e_{GS}$ at which $s_k (e)$ vanishes can be understood by a simple
energy balance argument. Flipping a cluster of $k$ spins among a
configuration with energy density $e$, one may gain a bulk energy
$E_b(k)\sim - k(e-e_{GS})$ and loose at least some surface energy $E_s (k)$.
As a consequence, $k$-stable configurations with energy densities
higher than $e_{GS}+E_s(k) /k$ should not
exist. For 1-D Ising disordered chain, $E_s (k)$ is equal to the
average magnitude of the smallest bond among $k$ ones and can be
easily computed. As expected $E_s (k) /k \sim e_k -e _{GS}$ for
large $k$. Let us emphasise that the previous argument implies that
there cannot be $\infty$-stable configurations (after the
thermodynamical limit has been taken) with energy densities differing
from the ground state one: $s_\infty (e \ne e_{GS}) = -\infty$. 

\section{Mean-field spin-glasses with infinite connectivity}
It is well known that the zero temperature limit of the complexity
equals the configurational entropy in infinite connectivity
spin-glasses\cite{beyond}. For instance in the case of the SK model,
\begin{equation}
\sigma ^{SK}= s_1 ^{SK} = \max _{z} \left\{ -\frac{z^2}{2} + \ln 
 2 \int_{-z}^{+\infty}Dy \right\} \simeq 0.1992 \label{sk0}
\end{equation}
with $Dy \equiv e^{-y^2/2}/\sqrt{2\pi }\; dy$ \cite{Bray1,0.19}. The
equivalence between TAP solutions at zero temperature and $1$-stable
configurations (IS) is a straightforward consequence of the vanishing
of couplings in the thermodynamical limit. The scaling of the
couplings $J_{ij}$ with size $N$ ensures that the excitation energy
$\Delta E_{i}$, corresponding to flipping spin $i$ remains finite in
the thermodynamical limit, e.g. $J_{ij} = O(1/\sqrt N)$ for the SK
model. As a consequence the excitation energy $\Delta E_{ij}=\Delta
E_{i}+\Delta E_{j}-2 J_{ij} S_i S_j$ corresponding to flipping both
spins $i$ and $j$ reduces to $\Delta E_{i}+\Delta E_{j}$ in the large
$N$ limit. For infinite connectivity models IS are also $k$-stable
configurations for every finite $k$. Therefore, $s_k(e) $ does not
depend on $k$ and $s_1 = \ldots =s_\infty = \sigma $.

\section{TAP solutions at zero temperature are $\infty$-stable 
configurations}
We now show that in presence of finite couplings, the degeneracy
between all $s_k$s is lifted since $k$-stable configurations are
not necessarily $k+1$-stable. In addition, we show that zero
temperature TAP solution are $\infty$-stable and not only
$1$-stable. To do so, let us first recall how TAP equations can be
derived for a one dimensional spin glass \cite{tapalbero}. We call
$l_{i}$ and $r_{i}$ respectively the effective magnetic fields due to
all the spins on the left and on the right of $i$ and $h_i$ the
external magnetic field acting on $S_i$. It is easy to write the
equation verified by $m_{i}$ in terms of $l_{i-1}$ and $r_{i-1}$,
\begin{equation}\label{mi}
m_{i}=\tanh \left[\beta h_i+\tanh^{-1}(\tanh (\beta l_{i-1})\tanh
(\beta J_{i-1})) +\tanh^{-1}(\tanh (\beta r_{i+1})\tanh (\beta
J_{i}))\right].
\end{equation}
This is the equation verified by the local magnetisation of a spin
$S_i$ interacting with two spins $S_{i-1}$, $S_{i+1}$ on which the
magnetic fields $l_{i-1}$ and $r_{i+1}$ act.  Focusing on two
neighbouring spins the equations verified by the two local
magnetisations are:
\begin{eqnarray}\label{meq1}
m_{i}&=&\frac{\tanh (\beta l_{i})+\tanh (\beta J_{i})\tanh (\beta
r_{i+1})} {1+\tanh (\beta l_{i})\tanh (\beta J_{i})\tanh (\beta
r_{i+1})}\\
\label{meq2}
m_{i+1}&=&\frac{\tanh (\beta r_{i+1})+\tanh (\beta J_{i}) \tanh (\beta
l_{i})} {1+\tanh (\beta l_{i})\tanh (\beta J_{i})\tanh (\beta
r_{i+1})}
\end{eqnarray}
These equations give $m_{i}$ and $m_{i+1}$ as a function of $l_{i}$
and $r_{i+1}$. Inverting (\ref{meq1}), (\ref{meq2}) one obtains
$l_{i}$ and $r_{i+1}$ as functions of $m_{i}$ and $m_{i+1}$.
Plugging $l_{i-1}=l_{i-1}(m_{i-1},m_{i})$ and $r_{i+1}=r_{i+1}
(m_{i},m_{i+1})$ into (\ref{mi}) the TAP equations for the local
magnetisations are established \cite{tapalbero}. 

To prove that TAP solutions are mapped into ground states in the zero
temperature limit let us focus on $k$ contiguous spins on the
chain. The TAP equations on these $k$ spins are by construction the
equations verified by the local magnetisations of the system of $k$
spins, as if it were isolated from the rest of the chain and there
were magnetic fields $l_{i_-}$, $r_{i_+}$ acting on the leftmost and
rightmost spins respectively. As a consequence in the zero temperature
limit, the $k$ local magnetisations tend towards the configuration of
$k$ spins which realizes the global minimum of the isolated system of
$k$ spins for any value of $l_{i_-}$ and $r_{i_+}$.  As $k$ can be
made arbitrarily large once $N$ has been sent to infinity, the
configurations corresponding to the zero temperature limit of TAP
solutions are $\infty$-stable. The argument can be straightforwardly
extended to Ising spin glasses on Cayley trees or on random graphs.

\section{Calculation for a mean-field model with finite couplings}
The results of the previous sections suggest that in finite
connectivity models $s_k$ is a non trivial function of $k$ which tends
towards the zero temperature limit of the complexity $\sigma$ for
large values of $k$. To corroborate this point, it is interesting to
study a model whose complexity $\sigma$ is non zero. Consider the
Hamiltonian
$H= - \sum_{i < j} J_{ij} S_i S_j -\sum_{i=1}^{N} L_{i}S_{i}S_{i+1}$ .
The first term 
is the
usual SK Hamiltonian: couplings $J_{ij}$ are independent Gaussian
random variables of zero mean and variance $J^2/N$. The one dimensional
Hamiltonian involves disordered interactions $L_{i}$ which are
independent random variables vanishing with probability $1-\alpha $ and
distributed according to an even distribution law $P(L)$ with
probability $\alpha$. This system smoothly interpolates between the SK
model ($\alpha =0$) and the one dimensional spin glass ($J=0$).
We calculate below the departures of $\sigma$ and $s_k$ from their
common value in the SK model to the first order in $\alpha$ only to avoid
useless tedious calculations.

To compute the complexity two approaches can be used.  The first one
consists in making the (annealed) sum over all locally stable TAP
solutions \cite{Bray1}. The other method relies on the computation of the
logarithm of the (average) $n^{th}$ power of the partition function
using Bray and Moore replica symmetry breaking scheme\cite{Bray2}.  As
$n$ goes to zero, the annealed complexity is
recovered\cite{Remi}.  Calculations are sketched in
Appendix A and B and provide the same average complexity $\sigma
(\alpha ) = \sigma ^{SK} + \alpha\; \gamma ^{(1)}  +O(\alpha ^2 )$ where  
\begin{equation}
\gamma ^{(1)} = -1+\left(\int_{-\infty}^{+\infty} dL P(L) 
\int _{-z^* - L/J}^{+\infty} Dy_1  Dy_2 \;\theta(y_1+y_2+2z^*
)\right)\bigg/ \left(\int_{- z^* }^{+\infty} Dy \right)^2
\ . \label{Ninf}
\end{equation}
and $z^* \simeq 0.506$ is the maximum of (\ref{sk0}). $\theta$ denotes the
Heaviside function.

We now turn to the computation of $s_k$.  The excitation energy
corresponding to $k$ spin flips depends on the SK couplings and the
initial configuration through the variables $x_i=\sum_{j (\ne
i)}J_{ij} S_i S_j$ only.  For large $N$, the $x_i$s become
independent of the spin configuration $\{S_i\}$ and can be written as
$x_i=y_i+z$, where $y_i$ and $z$ are independent Gaussian random
variables of zero means and variances $J^2$ and $J^2/N$
respectively\cite{Gross}. The annealed configurational entropy is defined
through
\begin{equation}\label{N1a}
e^{N s_1} = \left< \sum_{\{S_i\}}\prod_{i=1}^N\theta \left( 
L_{i-1}S_{i-1}S_i+L_{i}S_{i}S_{i+1}
+y_i+z\right)\right>_{\{L,y,z\}} \ .
\end{equation}
Once the average $\langle \cdot \rangle $ over disorder has been carried
out, spins may be traced over and (\ref{N1a})
may be evaluated for large $N$ by the method of steepest descent.
We find $s_1 (\alpha ) 
= \sigma ^{SK} + \alpha \; c_1 ^{(1)} + O(\alpha ^2)$ with
\begin{equation}\label{N2a}
c_1 ^{(1)} = -1+\left(\int_{-\infty}^{+\infty} dL P(L) 
\int _{-z^* -L/J}^{+\infty} Dy_1  Dy_2 
\right) \bigg/ \left(\int_{-z^* }^{+\infty} Dy \right)^2
\ ,
\end{equation}
giving $s_1 > \sigma $ as expected, compare (\ref{Ninf}) and
(\ref{N2a}) \cite{Dean}.  Repeating the same calculation for $k(\ge 2)$-stable
configurations, we have found that they are equally numerous to the
first order in $\alpha$: $s_k (\alpha ) = \sigma ^{SK} + \alpha \; c_k
^{(1)} + O(\alpha ^2)$ with $c_k ^{(1)}=\gamma ^{(1)}$ for any $k\ge 2$. This
result extends to higher orders in $\alpha$ as follows. Let us call
$c_k ^{(n)}$ the coefficient of $\alpha ^{n}$ in $s_k$. Contributions
to $c_k ^{(n)}$ come from disordered lattices of $L$ bonds involving
clusters of at most $n+1$ contiguous spins along the chain.  Clearly,
spin configurations on such lattices that are $n+1$-stable are for
ever stable. Therefore, $c_1 ^{(n)} > c_2 ^{(n)} > \ldots > c_{n+1}
^{(n)} = c_{k} ^{(n)}$ for any $k\ge n+1$. We have in particular
explicitely checked that $c_2 ^{(2)} > c_3 ^{(2)}$.

As a conclusion, we have found a mean-field model with non zero
complexity $\sigma$ and  $ s_1 > \sigma$. We have checked that $\sigma $
coincides with $s_\infty$ to the first
order in the density of $O(1)$ couplings.

\section{Conclusion}
The above examples have shown the differences between IS and PS even
in mean-field models provided that finite couplings are present. Our
results could also be extended to models with continuous degrees of
freedom, e.g. systems of $N$ interacting particles\cite{logx2}. In
this case IS are configurations stable against {\em infinitesimal}
moves of any subset of the $N$ particles, that is local minima of the
potential energy. $k$-stable local minima can be defined as IS stable
against {\em finite} moves of any subset of $k$ (or less than $k$)
particles. $\infty$-stable minima gather all configurations that can be
reached from each other by crossing finite barriers and are related to
PS.

From the equilibrium point of view, PS and thus complexity are the only
relevant concepts. The decomposition of the partition function put
forward by Stillinger based on IS though mathematically exact does not
seem to be thermodynamically founded\cite{logx1}.  However,
usual nucleation arguments imply that
no PS with free-energy $f\ne f_{eq}$ can live forever, that is 
$\sigma (f\ne f_{eq}) = -\infty$. We thus face the following
alternative. Either nucleation arguments break down for some reason 
and finite-dimensional glassy systems with full curves for $\sigma (f)$ 
may exist\cite{first,third}. Or they hold  and the KTW entropy
crisis scenario cannot be extended to realistic systems without taking
into account the notion of life time for PS (as was already emphasized
in \cite{Ktw}).  IS, besides their phenomenological
interest can be seen as a very valuable step
in that direction.

{\bf Acknowledgements:} We are very grateful to Marc M{\'e}zard for
very interesting and motivating discussions.

\section{Appendix A}
To first order in $\alpha$ spin $i$ has at most one neighbour $v=i\pm
1$ along the chain of $L$ bonds. As a consequence the TAP equations 
can be written as
\begin{equation}\label{appmeq}
m_{i}=\frac{t_{i}+t_{v}\tanh (\beta L)}
{1+t_{i}t_{v}\tanh (\beta L)} \ , \quad
\tanh^{-1}(t_{i})
=\beta \sum_{j (\ne i)} J_{ij}m_{j}-\beta ^{2}J^{2} (1-q)
m_{i}
\quad ,
\end{equation}
where $q=1/N\sum_{i=1}^{N}m_{i}^{2}$. To compute the
 annealed complexity we sum over all solutions of equations 
 (\ref{appmeq}) following \cite{Bray1}.
Introducing auxiliary variables one can perform the average over $J_{ij}$
 and using the same notation as \cite{Bray1}, we find 
\begin{eqnarray}\label{appN}
\sigma&=& \hbox{\rm extr}_{q,\lambda ,\Delta } \left\{ -\lambda q-
\Delta (1-q)-\Delta ^{2}/2\beta ^{2}J^{2}+\right. \\
&&\left. \frac{1}{N}
\ln \left<\int_{-1 }^{+1 }\prod _{i=1}^{N}
\frac{dt_{i}}{\beta J\sqrt{2\pi q}}\frac{1}{1-t_{i}^{2}}
\exp \left(-\frac{(\tanh ^{-1}t_{i}-\Delta m_{i})^{2}}{2\beta ^{2}J^{2}q}
+\lambda m_{i}^{2} \right)
\right>_{\{L \}}\right\}\quad .
\end{eqnarray}
Performing the average on $L_{i}$ at the first order in $\alpha$ and taking
the zero temperature limit we obtain the expression of $\gamma$.

\section{Appendix B}
To compute the $n^{th}$ moment of the partition function, we use the
replica trick and obtain $n$ coupled
chains. Defining the overlap $q^{ab}$ between replicas $a$ and $b$ 
and the $2^n\times 2^n$ transfer matrix,
\begin{equation}
{\cal M} (\vec S ,
\vec T ) = \left\{ 1 -\alpha +\alpha \int dL P(L) \exp \left[ \beta L 
\sum _a S ^a T ^a \right] \right\} \; \exp
\left[ \frac {\beta ^2 J^2 }{2} \sum _{a<b}
q^{ab} ( S^a S ^b + T ^a T ^b ) \right] \ ,
\end{equation}
the complexity may be computed within Bray and Moore two-group Ansatz
\cite{Bray2}: $ \sigma = \hbox{\rm extr}_{\{q \}} [- (\beta J )^2 \sum
_{a<b} (q^{ab})^2 /2 + \ln \Lambda ( \{ q\} )]$ where $\Lambda ( \{
q\} )$ is the largest eigenvalue of ${\cal M}$. The saddle-point
equation for $\{ q\}$ read $q^{ab}=\sum _{\vec S} S^a S^b [\Psi (\vec
S)]^2$. The normalised ground state wave function $\Psi$ obeys the two group
symmetry with breakpoint $m$\cite{Bray2}. $\Psi$ depends on $\vec S =
(S^1, \ldots , S^n)$ through $S_- = \sum _{a \le m} S^a$ and $S_+ =
\sum _{a > m} S^a$ only and can be written as $\Psi (S_- , S_+ ) = c .
\int dh_0 dh_1 \rho (h_0 , h_1 ) \exp ( \beta h_0 (S_+ +S_-) +
h_1 (S_+ -S_-) )$ where $c$ is a normalisation factor. In the zero
temperature limit, the effective fields $h_0, h_1$ are both of order
one and the overlaps $q^{ab}$ read $q_1=A-z/(\beta m J), q_2 = A -
w/(\beta m J)^2, q_3 = A + z/(\beta m J), A\to 1$ with the notations 
of \cite{Bray2}. The eigenvalue equation for $\Lambda ,
\Psi$ provides a linear equation for the effective field distribution
\begin{eqnarray}
&\lambda &\rho (h_0 , h_1 ) =  \int dL
P(L) \bigg\{ (1 -\alpha) \delta (L)  +\alpha 
P(L) \bigg\} \int Du Dv \int dh' _0 dh' _1  
\exp \bigg( h' _1 \omega _L (h' _0 ) \bigg)\nonumber \\
&& \times \rho \bigg( h' _0 -J u , h' _1 -u z - v \sqrt{2w-z^2} \bigg)
\; \delta \bigg( h_0 - \varphi _L (h' _0 ) \bigg)
\; \delta \bigg( h_1 - h' _1 
\eta _L (h' _0 ) \bigg) \ , \label{bord}
\end{eqnarray}
where $\lambda = \Lambda \exp(\beta J m z)$. 
$\omega _L (x)$ (respectively $\varphi _L (x)$, $\eta _L (x)$)
equals $-1$ (resp. $-L$, $0$) if $x<-|L|$; $0$ (resp. $x \; \hbox{\rm
sign}(L)$, $\hbox{\rm sign}(L)$) if $-|L|<x<|L|$ and $1$ (resp.
$L$, $0$) if $x>|L|$. The complexity $\sigma$ is found through
maximising $-z^2/2-w+\ln \lambda(z,w)$ over $z,w$ and can in principle
be computed for any
$\alpha$. The small $\alpha$ expansion of $\sigma$ is computed
from the expansion of $\rho$ around the SK distribution $\rho ^{SK} 
(h_0, h_1) =\delta (h_0) \delta (h_1)$ using (\ref{bord}).


\begin{thebibliography}{99}
%
\bibitem{Ktw}
T.R. Kirkpatrick, P.G. Wolynes, {\em Phys. Rev. A} {\bf
35}, 3072 (1987); {\em Phys. Rev. B} {\bf 36}, 8552 (1987).\\
T.R. Kirkpatrick, D. Thirumalai, {\em Phys. Rev. B} {\bf
36}, 5388 (1987); {\em Phys. Rev. B} {\bf 38}, 4881 (1988).

\bibitem{Angell}
M.D. Ediger, C.A. Angell, S.R. Nagel, {\em J. Phys. Chem.} {\bf 100},
13200 (1996). \\
C.A. Angell, {\em Physica D} {\bf 107}, 122 (1997).

\bibitem{first}
M. M{\'e}zard, G. Parisi, {\em Phys. Rev. Lett. }{\bf 82}, 747 (1998).

\bibitem{second}
S. Franz, M. Cardenas, G. Parisi {\em J.Phys. A} {\bf 31}, L163 (1998).

\bibitem{Stillinger}
F.H. Stillinger, T.A. Weber, {\em Science} {\bf 225}, 983 (1984).

\bibitem{sciort}
F. Sciortino, W. Kob, P. Tartaglia, {\em Phys. Rev. Lett.} {\bf 83}, 
3214 (1999).

\bibitem{felix}
A. Crisanti, F. Ritort, {\em Potential energy landscape of simple
p-spin models for glasses}, preprint cond-mat/9907499 (1999).

\bibitem{third}
B. Coluzzi, G. Parisi, P. Verrocchio, {\em Phys. Rev. Lett.}, 
{\bf 84}, 306 (2000).

\bibitem{groundstate}
D.A. Huse, D.S. Fisher, {\em J. Phys. A} {\bf 20}, L997 (1987).\\
A. Bovier, J. Frolich, {\em J. Stat. Phys.} {\bf 44}, 347 (1986).

\bibitem{tap}
D.J. Thouless, P.W. Anderson, R.G. Palmer, {\em Phil. Magazine} {\bf
35}, 593 (1977).

\bibitem{Luck}
J.M. Luck, {\em ``Syst{\`e}mes d{\'e}sordonn{\'e}s
unidimensionels''}, CEA, Collection Al{\'e}a-Saclay (1992).

\bibitem{Moore}
R. Ettelaie, M.A. Moore, {\em J. Physique Lettres} {\bf 46}, L893
(1985). 

\bibitem{Li}
T. Li, {\em Phys. Rev. B} {\bf 24}, 6579 (1981).

\bibitem{logx1}
F.H. Stillinger, {\em J. Chem. Phys} {\bf 88}, 7918 (1988).

\bibitem{logx2}
M. M{\'e}zard, {\em private communication}.

\bibitem{beyond}
M. M{\'e}zard, G. Parisi, M. Virasoro,
{\em ``Spin Glass Theory and Beyond'' }, World Scientific, Singapore (1987).

\bibitem{Bray1}
A.J. Bray, M.A. Moore, {\em J. Phys. C} {\bf 13}, L469 (1980).

\bibitem{0.19}
F. Tanaka, S.F. Edwards, {\em J. Phys. F} {\bf 13}, 2769 (1980).\\
C. De Dominicis, M. Gabay, T. Garel, H. Orland, {\em J. de Physique}
{\bf 41}, 923 (1980).

\bibitem{tapalbero}
K. Nakanishi, {\em Phys. Rev. B} {\bf 23}, 3514 (1981).\\
D.R. Bowman, K. Levin, {\em Phys. Rev. B} {\bf 25}, 3438 (1982).

\bibitem{Bray2}
A.J. Bray, M.A. Moore, {\em J. Phys. C} {\bf 13}, L907 (1980).

\bibitem{Remi}
R. Monasson, {\em Phys. Rev. Lett.} {\bf 75}, 2847 (1995).\\
G. Parisi, M. Potters, {\em Europhys. Lett.} {\bf 32}, 13 (1995).

\bibitem{Gross}
D.J. Gross, M. M{\'e}zard, {\em Nucl. Phys. B} {\bf 12}, 431 (1984).

\bibitem{Dean}
Recently, $s_1 (e)$ has been calculated for random thin graphs by
D. Dean; {\em Metastable states of spin glasses on random thin graphs}, 
preprint cond-mat/9912240 (1999).

\end{thebibliography}
\end{document}